# Deep Learning for Uplink Spectral Efficiency in Cell-Free Massive MIMO Systems

Le Ty Khanh[1,2], Viet Quoc Pham[1,2], Ha Hoang Kha[1,2], and Nguyen Minh Hoang[3]
[1] Ho Chi Minh City University of Technology (HCMUT), Vietnam
[2] Vietnam National University Ho Chi Minh City, Vietnam
[3] Saigon Institute of ICT (SaigonICT), Vietnam
Email: {khanhlety, vietpq09, nmhoang}@gmail.com, hhkha@hcmut.edu.vn

*Abstract*—In this paper, we introduce a Deep Neural Network (DNN) to maximize the Proportional Fairness (PF) of the Spectral Efficiency (SE) of uplinks in Cell-Free (CF) massive Multiple-Input Multiple-Output (MIMO) systems. The problem of maximizing the PF of the SE is a non-convex optimization problem in the design variables. We will develop a DNN which takes pilot sequences and large-scale fading coefficients of the users as inputs and produces the outputs of optimal transmit powers. By consisting of densely residual connections between layers, the proposed DNN can efficiently exploit the hierarchical features of the input and motivates the feed-forward nature of DNN architecture. Experimental results showed that, compared to the conventional iterative optimization algorithm, the proposed DNN has excessively lower computational complexity with the trade-off approximately only 1% loss in the sum-rate and the fairness performance. This demonstrated that our proposed DNN is reasonably suitable for real-time signal processing in CF massive MIMO systems.

## I. INTRODUCTION

In recent years, Deep Neural Networks (DNNs) have obtained exceptional achievements in a variety of applications, especially in image processing. In [1], the authors proposed a deep network architecture named as Residual Network (ResNet) which consists of shortcut connections between layers to counteract the degradation problem in training significant DNNs. Numerical simulations demonstrated that in image classification and recognition, ResNets can achieve excellent results with significant depth, e.g. 1000 layers. In [2], the authors introduced a dense convolutional network, namely DenseNet. By feeding feature maps of previous layers to all subsequent layers, the DenseNets not only mitigate the gradient vanishing but also excessively deduct the trainable parameters. Numerical results shown that, compared to other state-of-the-art methods in the task of image classification, especially the ResNets [1], the DenseNets can obtained lower error rates with less computational efforts. In image restoration, deep convolutional neural architectures usually lead to the long-term dependency problem in which layers close to the input have less impact on those close to the output, which degrades the performance of deep Convolutional Neural Networks (CNNs) in image denoising or super-resolution. Therefore, the authors in [3] introduced a very deep architecture named as memory network (MemNet) which consists of memory blocks to deal with the long-term dependency challenge. Numerical results demonstrated that the MemNets not only correctly recovered noisy images but also provided considerably sharper images compared to other advanced techniques, such as Super-Resolution CNN (SRCNN) [4], Very Deep Super-Resolution (VDSR) [5], and denoising CNNs (DnCNNs) [6]. Also, the authors in [7] proposed an innovative Residual Dense Network (RDN) which effectively exploits hierarchical features of low-resolution images in the input to achieve considerably high performance in the task of image super-resolution. Experiments shown that the RDNs provided images with higher resolution than those predicted by others advanced methods, such as MemNet [3] and SRDenseNet [8].

Recently, Deep Learning (DL) has been applying into Multiple-Input Multiple-Output (MIMO) wireless communication systems to combat the time-consuming challenge in signal processing. Regarding the task of detection in MIMO systems, in [9], [10], the authors introduced a deep neural architecture called DetNeT, i.e., detection network, which is motivated from unfolding the Gradient Projection (GP) method. Numerical results showed that the DetNet can provide similar accuracy as another advanced detector, e.g. Approximate Message Passing (AMP), however, with significantly lower complexity and without the requirement of knowledge of the noise variance. Also, the DetNeT demonstrated its robustness against ill conditioned channels. Regarding the task of sum-rate maximization, in [11], the authors used a deep neural architecture to address the problem of time consuming in Signal Processing (SP) of communication systems, which makes the SP problems feasible to the real-time deployment. Experimental results in [11] shown that the deep network approach requires excessively lower computational time, i.e., orders of magnitude, than the Weighted Minimum Mean Square Error (WMMSE) algorithm. In [12], the authors introduced the PowerNet which jointly optimizes the pilot and data powers to maximize the sum-rate of Multi-cell (MC) massive MIMO systems. Experiment results demonstrated that the PowerNet is suitable for real-time optimization in MC massive MIMO networks. Recently, DL has also been applied to Cell-Free (CF) mmWave massive MIMO systems. Given that CF mmWave massive MIMO networks can inherit advantages from both of the CF massive [13], [14] and the mmWave MIMO networks, the downside is the highly computational complexity due to the deployment of a great number of antennas and large bandwidth. Therefore, in [15], the authors proposed the Fast and Flexible Denoising convolutional Neural Network (FFDNet) for the task of channel estimation in CF mmWave massive MIMO systems. Experimental results shown that, with only one training model, the FFDNet not only can provide favorable Normalized Mean Squared Error (NMSE) at different noise levels but also outperforms the classical Minimum Mean Square Error (MMSE) estimator.

In this paper, we will introduce a deep residual dense network for power allocation, named as PowerRDN, to address the Proportional Fairness (PF) maximization problem in uplinks of CF massive MIMO systems. In recent years, both of CF massive MIMO networks and applications of deep neural architectures in communication systems have received a growing concern from the academic society [9], [10], [13], [14]. However, to the best of our knowledge, there has been no prior work on maximizing the PF problem in CF massive MIMO systems by using densely DNNs. Therefore, we will briefly describe an iterative algorithm based on the generalized eigenvalue problem and the Gradient Projection (GP) method to solve the coupled non-convex PF problem. Optimal solutions from the iterative algorithm then will be used to generate training data for the proposed PowerRDN. We demonstrate that the PowerRDN can reduce computational time to orders of magnitude with only 1-2% loss in the performance, e.g., the sum-rate, compared to the iterative optimization approach.

*Outline:* The remainder of this paper is organized as follows. In Section II, the CF massive MIMO system and the PF optimization problem will be modeled and formulated, respectively. In Section III, we will present an iterative algorithm resolving the PF problem. In Section IV, the architecture of our proposed PowerRDN will be presented in detail. In Section V, numerical results will be provided to evaluate the performance of the proposed network. Discussion and conclusion will be given in Section VI.

*Notation:* Matrices and column vectors are respectively represented by upper and lower boldface letters. The superscript $(.)^T$ and $(.)^H$ are the transpose and the conjugate-transpose operators, respectively while $\mathbf{E}[.]$ is the expectation operator. $\|.\|$ represents the Euclidean norm. $\odot$ and $\oslash$ represent the Hadamard product and division, respectively. With zero mean and variance $\sigma^2$, circularly symmetric complex Gaussian Random Variable (RV) $v$ is denoted as $v_1 \sim \mathcal{CN}(0, \sigma^2)$, while real value Gaussian RV $v_2$ is presented as $v_2 \sim \mathcal{N}(0, \sigma^2)$.

## II. SYSTEM MODEL

We consider the CF massive MIMO system in which $M$ single-antenna Access Points (APs) simultaneously serves $K$ single-antenna users [13]. The APs are communicated to a central processing unit (CPU) via a perfect backhaul network. In this paper, we focus on uplink training and uplink payload data transmission.

In the uplink training phase, channel coefficients from the users to the APs are estimated. Denote channel coefficient between AP $m$ and user $k$ by [13].

$$h_{mk} = \sqrt{\beta_{mk}} \tilde{h}_{mk} \qquad (1)$$

where $\tilde{h}_{mk} \sim \mathcal{CN}(0,1)$ denotes the small-scale fading while $\beta_{mk}$ captures the large-scale fading. Concerning channel estimation, let the pilot sequence of $\tau$ symbols transmitted from user $k$ be denoted as $\sqrt{\tau}\boldsymbol{\varphi}_k \in \mathbb{C}^{\tau \times 1}$. These pilot sequences are chosen from a predetermined set [13]. The received pilot vector at AP $m$ is given as:

$$\boldsymbol{s}_{\mathrm{p}m} = \sum_{k=1}^{K} \sqrt{\rho_{\mathrm{p}}} \sqrt{\tau} h_{mk} \boldsymbol{\varphi}_k + \boldsymbol{n}_{\mathrm{p}m} \qquad (2)$$

where $\rho_{\mathrm{p}}$ is the normalized signal-to-noise ratio (SNR) of each pilot symbol. $\boldsymbol{n}_{\mathrm{p}m} \sim \mathcal{CN}(0;1)$ is a vector of additive noise at AP $m$. The channel coefficients are estimated at the APs by applying the MMSE estimator to the received pilot sequences at the APs [13]. AP $m$ uses $\boldsymbol{\varphi}_k$ to decorrelate the received signal.

$$\bar{s}_{\mathrm{p}m} = \boldsymbol{\varphi}_k^H \boldsymbol{s}_{\mathrm{p}m} = \sum_{k=1}^{K} \sqrt{\rho_{\mathrm{p}}} \sqrt{\tau} h_{mk} \boldsymbol{\varphi}_k^H \boldsymbol{\varphi}_k + \boldsymbol{\varphi}_k^H \boldsymbol{n}_{\mathrm{p}m} \qquad (3)$$

By using MMSE estimator, channel estimate $\hat{h}_{mk}$ is given as $\hat{h}_{mk} = \vartheta_{mk} \bar{s}_{\mathrm{p}m}$ where:

$$\vartheta_{mk} \triangleq \frac{\mathbf{E}\left[\bar{s}_{\mathrm{p}m}^* h_{mk}\right]}{\mathbf{E}\left[\left|\bar{s}_{\mathrm{p}m}\right|^2\right]} = \frac{\sqrt{\tau\rho_{\mathrm{p}}}\beta_{mk}}{\tau\rho_{\mathrm{p}} \sum_{i=1}^{K} \beta_{mi} \left|\boldsymbol{\varphi}_k^H \boldsymbol{\varphi}_i\right|^2 + 1} \qquad (4)$$

In uplink payload data transmission, let $x_k$ with $\mathbf{E}\left[\left|x_k\right|^2\right] = 1$ be a symbol transmitted from user $k$. $p_k$, $0 \le p_k \le 1$ is the normalized transmit power coefficient of user $k$. The received signal at AP $m$ is then:

$$s_m = \sum_{i=1}^{K} h_{mi} \sqrt{\rho} \sqrt{p_i} x_i + n_m \qquad (5)$$

where $\rho$ denotes the normalized uplink signal-to-noise ratio (SNR). $n_m \sim \mathcal{CN}(0,1)$ stands for the additive white Gaussian noise at AP $m$. Before transmitting the received signal to the CPU, AP $m$ multiplies the received signal with the conjugate of the channel coefficient from user $k$ to detect the transmitted symbol from user $k$. Also, in [13], [16], the CPU multiplies signal transmitted from the APs with receiver filter coefficients, $t_k = [t_{1k},...,t_{Mk}]^T, \|t_k\| = 1$, to improve user achievable rates. Therefore, the received superposition signal at the CPU can be given by:

$$s_k = \sum_{m=1}^{M}\sum_{i=1}^{K} t_{mk} \hat{h}_{mk}^* h_{mi} \sqrt{\rho} \sqrt{p_i} x_i + \sum_{m=1}^{M} t_{mk} \hat{h}_{mk}^* n_m \quad (6)$$

Since the CPU only deploys knowledge of channel statistics to detect user $k$'s information, this signal can be formulated as a sum of the desired and the beamforming uncertainty signals. By using the channel hardening property [13], the achievable rate of user $k$ in case of uncorrelated Gaussian noise is given by Eq. (7) at the bottom of this page.

$$R_k = \log_2\left(1 + \frac{\rho p_k \left(\sum_{m \in \mathcal{M}} t_{mk} \xi_{mk}\right)^2}{\sum_{i \in \mathcal{K}} \rho p_i \left(\sum_{m \in \mathcal{M}} t_{mk}^2 \xi_{mk} \beta_{mi}\right) + \sum_{i \in \mathcal{K} \setminus k} \rho p_i |\varphi_k^H \varphi_i|^2 \left(\sum_{m \in \mathcal{M}} t_{mk} \xi_{mk} \frac{\beta_{mi}}{\beta_{mk}}\right)^2 + \sum_{m \in \mathcal{M}} t_{mk}^2 \xi_{mk}}\right) \quad (7)$$

In this paper, we deploy the proportional fairness utility function [17], [18] to strike a balance between the achievable sum-rate and the fairness among users. The problem of proportional fairness maximization in the CF massive MIMO network can be mathematically formulated as:

$$\mathcal{P}_1: \quad \max_{t,p} f(t,p) \triangleq \sum_{k=1}^{K} \log_2(R_k) \quad (8a)$$

$$\text{s.t.} \quad \|t_k\| = 1, \quad k \in \mathcal{K} \quad (8b)$$

$$0 \leq p_k \leq 1 \quad (8c)$$

where we have defined $t \triangleq (t_k)_{k \in \mathcal{K}}$, $p \triangleq (p_k)_{k \in \mathcal{K}}$. Problem $\mathcal{P}_1$ is a non-linear and non-convex maximization problem due to the coupling between the receiver filter coefficients $t$ and the power variables $p$.

### III. ITERATIVE OPTIMIZATION ALGORITHM

To train the DDNs, the sets of the channel conditions and the corresponding optimal filters and power allocation to Problem $\mathcal{P}_1$ are obtained. To that end, in this section, we summary an iterative algorithm in [19] based on a gradient projection approach to solve Problem $\mathcal{P}_1$. A local optimal solution to Problem $\mathcal{P}_1$ can be achieved by alternatively solving the maximization problem with respect to the receiver filter coefficients while power variables are fixed and vice versa. The updated receiver filter coefficients are then fixed to update values of power variables for the next iteration.

#### A. Receiver Filter Coefficient Design

We define $\xi_k \triangleq [\xi_{1k},...,\xi_{Mk}]^T$, $\beta_k = [\beta_{mk},...,\beta_{Mk}]^T$, $\zeta_{ki} = |\varphi_k^H \varphi_i|^2 \xi_k \odot \beta_i \oslash \beta_k$, $\mathbf{Y}_{ki} = \text{diag}(\xi_k \odot \beta_i)$, and $\Xi_k = \text{diag}(\xi_k)$. The achievable rate of user $k$ can be restated as:

$$R_k = \log_2\left(1 + \frac{t_k^H (\rho p_k \xi_k \xi_k^H) t_k}{t_k^H \left(\sum_{i \in \mathcal{K}} \rho p_i \mathbf{Y}_{ki} + \sum_{i \in \mathcal{K} \setminus k} \rho p_i \zeta_{ki} \zeta_{ki}^H + \Xi_k\right) t_k}\right) \quad (9)$$

while keeping power variables as fixed values, the achievable rate of user $k$ only depends on its own receiver filter coefficients $t_k$ and is independent of receiver filter coefficients of other users. Therefore, Problem $\mathcal{P}_1$ can be solved by individually maximizing the achievable rate of the users [16], [20], [21]. Optimal $t_k$ can be obtained by solving the following problem:

$$\mathcal{P}_2: \quad \max_{t_k} \quad \text{SINR}_k \quad (10a)$$

$$\text{s.t.} \quad \|t_k\| = 1, \quad k \in \mathcal{K} \quad (10b)$$

where

$$\text{SINR}_k = \frac{t_k^H (\rho p_k \xi_k \xi_k^H) t_k}{t_k^H \left(\sum_{i \in \mathcal{K}} \rho p_i \mathbf{Y}_{ki} + \sum_{i \in \mathcal{K} \setminus k} \rho p_i \zeta_{ki} \zeta_{ki}^H + \Xi_k\right) t_k} \quad (11)$$

Let $\mathbf{D}_k = \sum_{i \in \mathcal{K}} \rho p_i \mathbf{Y}_{ki} + \sum_{i \in \mathcal{K} \setminus k} \rho p_i \zeta_{ki} \zeta_{ki}^H + \Xi_k$ and $\mathbf{N}_k = \rho p_k \xi_k \xi_k^H$. Problem $\mathcal{P}_2$ is a generalized eigenvalue problem [13], [19], [22] of the matrix pair $\mathbf{N}_k$ and $\mathbf{D}_k$. Therefore, optimal $t_k$ is the normalized eigenvector corresponding to the maximum generalized eigenvalue.

#### B. Power Allocation

By denoting $\alpha_k = \rho t_k^H \xi_k \xi_k^H t_k$, $\eta_{ki} = \rho t_k^H \zeta_{ki} \zeta_{ki}^H t_k$, $\chi_{ki} = \rho t_k^H \mathbf{Y}_{ki} t_k$, and $\delta_k = t_k^H \Xi_k t_k$, $\text{SINR}_k$ can be rewritten as:

$$\mathrm{SINR}_k = \frac{\alpha_k p_k}{\sum_{i \in \mathcal{K}} \chi_{ki} p_i + \sum_{i \in \mathcal{K} \setminus k} \eta_{ki} p_i + \delta_k} \quad (12)$$

By treating the receiver filter coefficients as fixed, Problem $\mathcal{P}_1$ can be rewritten as:

$$\mathcal{P}_3: \quad \max_p \quad f_{t_k^*}(p) \triangleq \sum_{k \in \mathcal{K}} \log_2(R_k) \quad (13a)$$

$$\text{s.t.} \quad 0 \leq p_k \leq 1, \quad k \in \mathcal{K} \quad (13b)$$

Then, the local optimal solution to Problem $\mathcal{P}_3$ can be achieved by using a Gradient Projection (GP) iterative algorithm [18], [19]. The GP approach is summarized in the following steps:

- Updating the power variables: $p_k' = k p_k + \varrho \nabla_{p_k} f_{t^*}(p)$, where $\varrho$ is the step size. $\nabla_{p_k} f_{t^*}(p)$ is the gradient of the objective function in Problem $\mathcal{P}_3$ with respect to variable $p_k$.
- Projecting the updated power variables, $(p_k')_{k \in \mathcal{K}}$, onto the feasible region, i.e. [0; 1], to ensure the satisfaction of the power constrain (13b). Let $(\bar{p}_k)_{k \in \mathcal{K}}$ denote the projected power variables.
- Applying Armijo's rule with provable convergence to the projected power variables [23], [24] to find the initial values of the power variables in the next iteration.

The steps resolving Problem $\mathcal{P}_1$ which are successively repeated until the algorithm reaches the tolerance error $\varepsilon$ are summarized in Algorithm 1.

---

**Algorithm 1**: Algorithm for Problem $\mathcal{P}_1$

**Initialize**:
- Set $\kappa = 1$, $search = \text{true}$, $\varepsilon = 10^{-3}$.
- Choose initial variables: $\varphi_k, p_k^{(1)}$, $\forall k$.

**Repeat**
1. Solve Problem $\mathcal{P}_2$ with fixed $p_k^{(\kappa)}$ to obtain $t^{*(\kappa+1)}$.
2. Solve Problem $\mathcal{P}_3$ with fixed $t_k^{*(\kappa+1)}$ to obtain $p^{(\kappa+1)}$.
3. Considering terminating condition

If $\dfrac{f\left(t^{(\kappa+1)}, p^{(\kappa+1)}\right) - f\left(t^{(\kappa)}, p^{(\kappa)}\right)}{f\left(t^{(\kappa)}, p^{(\kappa)}\right)} \leq \varepsilon$ then

$search = \text{false}$

**Else**

$\kappa = \kappa + 1$

**End if**

**Until** $search = \text{false}$

**Output**: $t^{(\kappa+1)}, p^{(\kappa+1)}$

---

## IV. PROPOSED DEEP LEARNING APPROACH

Regarding Problem $\mathcal{P}_1$, given that the maximization problem has two variables, i.e., receiver filter coefficients $t$ and power coefficients $p$, one variable can be found if we have knowledge about the other variable. In other words, if we know optimal values of receiver filter coefficients, we can find the optimal power coefficients, and vice versa. If optimal power coefficients are known, we can obtain the optimal values of the receiver filter coefficients. Therefore, it is sufficient to train a network performing a regression task such that outputs of the network are the optimal receiver filter coefficients or the optimal powers values. Technically speaking, in the regression task, the smaller size of the output matrix, the better performance of the model. Therefore, instead of using a neural network to predict optimal receiver filter coefficients, anticipating optimal power coefficients will result in better solutions to Problem $\mathcal{P}_1$ as the number of the power coefficients is much less than the number of the receiver filter ones. In the following subsection, we will present our proposed deep neural architecture, namely PowerRDN, which is used for predicting optimal powers.

### A. Architecture

The architecture of our PowerRDN is shown in Fig. 1. Let $\boldsymbol{\mu} \in \mathbb{N}^K, \mu_k \in [1; \tau], \tau < K$ be a vector whose elements are integer numbers corresponding to the orders of the users' pilot sequences in the predetermined pilot set. Particularly, if two users $i$ and $j$, $i, j \in \mathcal{K}$, use the same pilot sequence, e.g., the $\ell$-th pilot sequence in set $\mathcal{S}$, during the training phase, then $\mu_i = \mu_j = \ell$. We name $\boldsymbol{\mu}$ as the Pilot Order Vector (POV). Let:

$$\mathbf{I}_{\mathrm{PF}} = \begin{bmatrix} \mu_1 & \boldsymbol{\beta}_1^T \\ \mu_2 & \boldsymbol{\beta}_2^T \\ \vdots & \vdots \\ \mu_K & \boldsymbol{\beta}_K^T \end{bmatrix} \in \mathbb{R}^{K \times (M+1)} \quad (14)$$

The first column of $\mathbf{I}_{\mathrm{PF}}$ is the POV vector. The n-th column of $\mathbf{I}_{\mathrm{PF}}$, $n = 2, \ldots, M+1$, contains large-scale fading coefficients from $K$ users to the (n-1)-th AP. Matrix $\mathbf{I}_{\mathrm{PF}}$ will be taken as the input for our deep neural network. Let $\boldsymbol{p}^* \in \mathbb{R}^K, p_k \in [0; 1], k \in \mathcal{K}$ be the output of the network. Vector $\boldsymbol{p}^*$ contains the predicted optimal powers of the users. The network architecture consists of three parts: Feature Extraction Layer (FEL), Residual Dense Block (RDB) [7], and Feature Reinforcement Layer (FRL). For simplification, we will omit the biases when describing the operation of the network by equations.

- Feature extraction layer: The FEL consists of a convolution operator followed by a tanh activation function as follows:

$$B_0 = \mathbf{A}_{\mathrm{FEL}}\left[\mathbf{C}_{\mathrm{FEL}}(\mathbf{I}_{\mathrm{PF}})\right] \quad (15)$$

$\mathbf{A}_{\mathrm{FEL}}$ is an activation function. $\mathbf{C}_{\mathrm{FEL}}$ is a $1 \times (M+1)$ convolution operator with the growth rate G [2], i.e., G feature maps.

- Residual Dense Block (RDB): The model consists of an RDB which is motivated from [7]. The block

consists of $L$ dense connected layers [2], a feature fusion, and a residual learning. The first layer is a combination of a 3×3 convolution and an activation function. The next (L-1) layers have the same architecture, which is sequentially constituted by a concatenation, a 3×3 convolution, and an activation function. The output of the $\ell$-th layer of the RDB can be written as:

$$B_\ell = \mathbf{A}_{B,\ell}\left[\mathbf{C}_{B,\ell}\left([B_0, B_1, \ldots, B_{\ell-1}]\right)\right] \quad (16)$$

The mechanism of forwarding feature maps of previous layers to all subsequent layers motivates the preservation of the feed-forward natures and the extraction of local dense features. This mechanism combined with residual learning was named as contiguous memory mechanism [7]. Since the more number of dense connected layers, the more number of concatenated feature maps at the end of the block, which is challenging when training a very deep neural network. Therefore, we used a concatenation followed by a 1x1 convolution with G feature maps in the last layer of the RDB, the (L+1)-th layer, as a fusion over feature channels to reduce the dimensionality of the feature maps. 1x1 convolutions were introduced in [25] and then were widely used in other works, such as in MemNeT [3] and RDN [7]. Let $B_G$ be the output of the (L+1)-th layer. The output of the RDB denoted as $B_F$ is achieved by:

$$B_F = B_0 + B_G \quad (17)$$

- Feature Reinforcement Layer (FRL): the FRL consists of a 3×3 convolution with only a single feature map followed by an activation function. The outputs of the FRL are a vector consisting of predicted transmit powers of the users, $\mathbf{p}^*$.

The Euclidean norm, i.e., the Root-Mean-Square Error (RMSE), is used to evaluate the performance of the trained network as follows:

$$\text{RMSE} = \left\|\mathbf{p} - \mathbf{p}^*\right\| \quad (18)$$

$\mathbf{p}$ and $\mathbf{p}^*$ is the true and predicted optimal transmit powers.

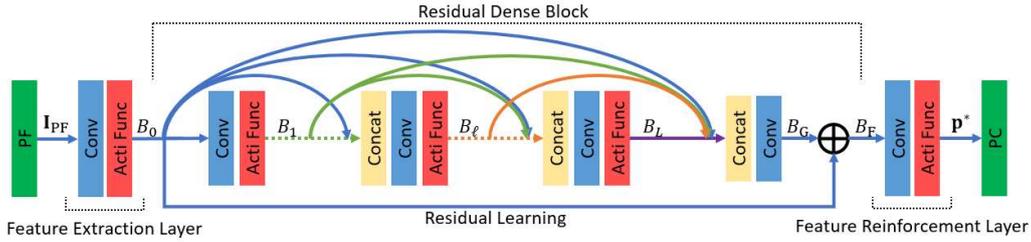

Figure 1. Our proposed deep neural network.

### B. Discussion

Given that our proposed PowerRDN is motivated from previous works, especially Residual Dense Network (RDN) in [7]. As compared to the RDN in [7] which consists of some RDBs with Global Residual Learning (GRL), our PowerRDN has only one RDB. The omission helps simplify the model, hence reducing computational time, as the task of powers prediction in this paper has less number of output elements than it was in the task of resolution enhancement in [7]. Also, with the FRL, we prefer using the sigmoid and tanh functions to deploying the ReLU function as activation functions. This can be explained as the ReLU layer outputs non-negative values without upper bound, which is suitable for predicting value of pixels in the range of [0; 255]. Meanwhile, the sigmoid and the tanh functions are nonlinear and bounded. This makes the use of the tanh function or the sigmoid function as activation layers is much more suitable to the task of power prediction than the use of ReLU function.

### C. Complexity Analysis

In this subsection, we will evaluate complexity of the PowerRDN. The numbers of trainable parameters in each component of the PowerRDN are given as follows:

- The FEL consists of $G \times 1 \times (M+1)$ weights and $G$ bias.
- The RDB consists of $L$ dense connected layers and a feature fusion. The $\ell$-th layer in the RDB consists of $G \times 3 \times 3 \times (\ell * G)$ weights and $G$ bias. Therefore the number of trainable parameters of the first $L$ layers is $G \times 3 \times 3 \left(\dfrac{L*(L+1)}{2} \times G\right)$ weights and $L \times G$ bias. The feature fusion consists of $G \times 1 \times 1 \times ((L+1) \times G)$ weights and $G$ bias.
- The FRL consists of $3 \times 3 \times G$ weights and single bias.

The total number of trainable parameters in the PowerRDN are summarized in Table I.

TABLE I.  NUMBER OF TRAINABLE PARAMETERS

| Layer | No. of weights | No. bias |
|---|---|---|
| FEL | $G \times (M+1)$ | $G$ |
| RDB (L layers) | $9\dfrac{L(L+1)}{2}G^2$ | $L \times G$ |
| RDB (feature fusion) | $(L+1)G^2$ | $G$ |
| FRL | $9G$ | 1 |

## D. Data Set and Training Process

For each setup of CF Massive MIMO systems, e.g., 80 APs and 20 users, Algorithm 1 is deployed to generate 12,000 samples for training and testing the PowerRDN. Particularly, the training set contains 11,000 samples in which 10,000 realizations are used for training and 1,000 samples are used for validation during training. The PowerRDN is trained through 40 epochs with the mini-batch size of 128 samples. The initial learning rate is $10^{-4}$. The learning rate drop factor is chosen as 0.1 for every 20 epochs. Adam optimization [26] is used to train the PowerRDN.

## V. NUMERICAL RESULTS

The APs and the users are randomly distributed in a $D \times D$ square. The wrapped around topology [13] was utilized to simulate a wide coverage area without boundary. In all experiments, we denote the path loss between user $k$ and AP $m$ as $PL_{mk}$ which is modeled as in [13]:

$$PL_{mk} = \begin{cases} -L - 35\log_{10}(d_{mk}), & d_1 < d_{mk} \\ -L - 15\log_{10}(d_1) - 20\log_{10}(d_{mk}), & d_0 < d_{mk} \leq d_1 \\ -L - 15\log_{10}(d_1) - 20\log_{10}(d_0), & d_{mk} \leq d_0 \end{cases}$$
(19)

where $L = 140.7 \text{dB}$, which is modeled as in [13], [14]. We consider uncorrelated shadowing with the large-scale fading coefficient in Equation (1) modeled as in [13]:

$$\beta_{mk} = 10^{\frac{PL_{mk}}{10}} 10^{\frac{\sigma_{sh} v_{mk}}{10}}$$
(20)

In the equation above, $PL_{mk}$ is the path loss. $10^{\frac{\sigma_{sh} v_{mk}}{10}}$ model shadow fading with standard deviation $\sigma_{sh} = 8\text{dB}$ and $v_{mk} \sim \mathcal{N}(0,1)$. Noise power $\rho_n = -92\text{dBm}$, which is modeled as [13], [14], [27]. Normalized pilot and data transmit powers are given as:

$$\rho_p = \frac{\bar{\rho}_p}{\rho_n}, \quad \rho = \frac{\bar{\rho}}{\rho_n}$$
(21)

where $\bar{\rho}_p$ and $\bar{\rho}$ are pilot and transmit powers. The overhead of the UL channel estimation had been taken into account in all simulation results as

$$R_{net,k} = \frac{1 - \tau/\tau_c}{2} R_k$$
(22)

where $\tau_c = 200$ and $\tau = 10$ samples. Other simulation parameters are given in Table II. The simulations, including the training phase of the PowerRDN, are implemented in MATLAB on a laptop with a core i5 CPU, 2.5GHz.

TABLE II. SIMULATION PARAMETERS

| Parameters | Values |
|---|---|
| $D, d_1, d_0$ | 1000m, 50m, 10m |
| B | 20MHz |
| W | 9dB |
| $\bar{\rho}_p, \bar{\rho}$ | 200mW |
| $\varepsilon$ | 0.001 |

*Example 1*: In this example, we will evaluate the sum-rates achieved by Algorithm 1 and the PowerRDN with 20 users. The sum-rates are shown in Fig. 2. As can be viewed in the figure that the sum-rates achieved by the PowerRDN are nearly identical to those obtained by Algorithm 1. Particularly, with 80 APs, the PowerRDN provides the sum-rate at 19.1422 [bits/s/Hz] while it is 19.2592 [bits/s/Hz] in case of Algorithm 1. The ratio of the sum-rate achieved by the PowerRDN to the sum-rate obtained Algorithm 1 is approximately 99.39%.

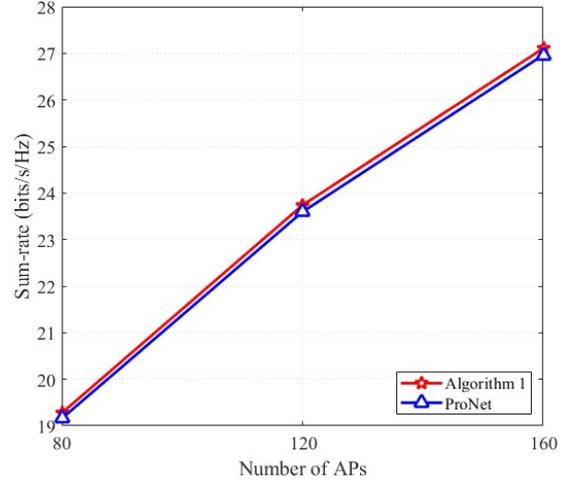

Figure 2. Sum rate of 20 users, 1000 samples.

*Example 2*: In this example, we adopt Jain's index:

$$J = \frac{\left(\sum_{k \in \mathcal{K}} R_{net,k}\right)^2}{K \sum_{k \in \mathcal{K}} R_{net,k}^2}$$
(23)

to evaluate the PF of the SE maximization achieved by Algorithm 1 and the PowerRDN. Particularly, $J = \frac{1}{K}$ and $J = 1$ are corresponding to the least and most PF of the SE, respectively. Regarding the CF Massive MIMO systems with 20 users, Jain's indices achieved by the PowerRDN and Algorithm 1 are shown in Fig. 3. As it was shown in the figure that, the Jain's indices provided by the PowerRDN are approximately 98.5% of those given be Algorithm 1. Particularly, with 80 APs, the index provided by the PowerRDN is 0.95496 while it was 0.96954 in case of Algorithm 1.

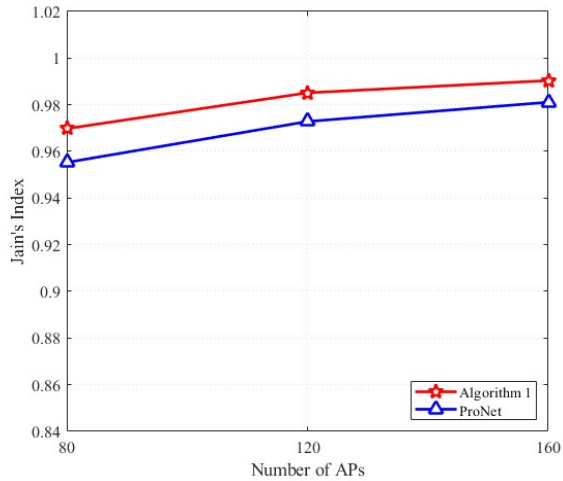

Figure 3. Jain's index of 20 users with 1000 samples.

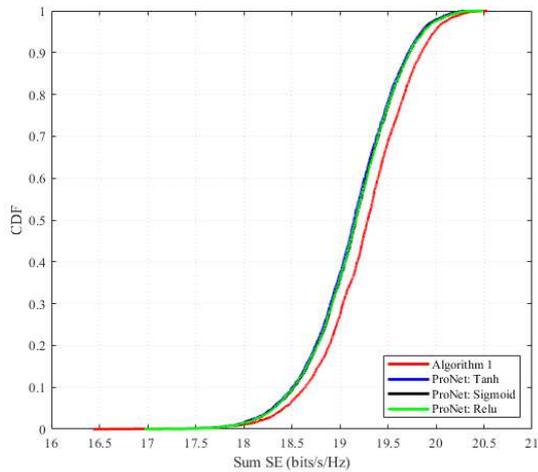

Figure 4. CDF of Sum SE of 20 users.

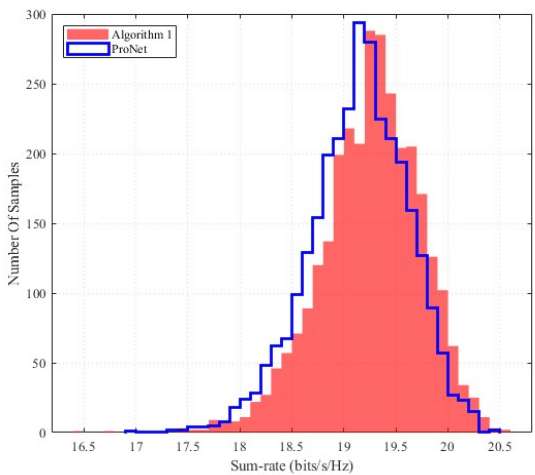

Figure 5. Histogram Sum-rate of 20 users, 3000 samples.

*Example 3*: In this example, we will evaluate the distribution of the sum-rates achieved by both approaches, i.e. the PowerRDN and Algorithm 1. Fig. 4 and Fig. 5 show the Cumulative Distribution Functions (CDFs) and the histogram of the sum-rates in case of 20 users. In Fig. 4, the curve with the legend of ReLU denotes the PowerRDN model with all activation functions are ReLU except the $\mathbf{A}_{FRL}$ is the sigmoid function. As it can be seen from the figures that the sum-rate distribution obtained by the PowerRDN is approximate as it is in case of using Algorithm 1 with only a minor gap in the SE performance. Numerical results are summarized in Table III. Fig. 5 shows the histogram of the sum rates achieved by the PowerRDN model with all activation functions are the tanh function. As observed from Fig. 5, the sum-rate distribution obtained by both approaches are quite similar.

TABLE III. COMPARISON

| Characteristic | PowerRDN (P) | Algorithm 1 (A) | P/A |
|---|---|---|---|
| Sum rate (bit/s/Hz) | 19.1422 | 19.2592 | 99.39% |
| Jain's index | 0.95496 | 0.96954 | 98.50% |
| Time (s) | 0.12081 | 14.1431 | 0.85% |

## VI. CONCLUSION

In this paper, we have introduced a deep neural network named as PowerRDN to address the PF of the SE maximization problem of uplinks in CF massive MIMO systems. Due to the coupling of the receiver filter coefficients and transmit power variables, the objective problem is non-convex. An iterative algorithm has been adopted to solve the maximization problem and to generate the training data set for the PowerRDN. The inputs of the PowerRDN is a matrix constituted from pilot sequences and large-scale fading coefficients of the users. Optimal transmit powers will be predicted by the PowerRDN. Based on the predicted powers, optimal receiver filter can be achieved by solving the generalized eigenvalue. Experimental results demonstrated that our PowerRDN is favorably suitable for real-time signal processing in CF assive MIMO networks as the PowerRDN has extremely lower computational time than the iterative optimization algorithm with only approximately 1% trade-off in the SE and the fairness performances.

## CONFLICT OF INTEREST

The authors declare no conflict of interest.

## AUTHOR CONTRIBUTIONS

L. T. Khanh and V. Q. Pham conducted the research and the numerical simulations; V. Q. Pham and H. H. Kha analyzed the data; L. T. Khanh and V. Q. Pham wrote the paper; N. M. Hoang and H. H. Kha edited the paper; all authors had approved the final version.

## ACKNOWLEDGMENT

This research is funded by Vietnam National Foundation for Science and Technology Development (NAFOSTED) under grant number 102.04-2017.308.

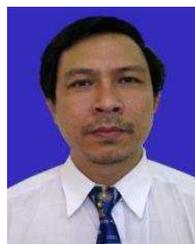

**Le Ty Khanh** was born in Phu Yen, Vietnam. He received the B.Eng. degree from Da Nang University in 1987, the second B.Eng. degree from Ho Chi Minh City University of Technology in 1996, and M. Eng. degree from Da Nang University in 2009. He is working at Phu Yen Department of Information and Communications. He is currently a Ph.D. candidate at the Faculty of Electrical and Electronics Engineering, Ho Chi Minh City University of Technology, Vietnam. His research interests are in digital signal processing and wireless communications.

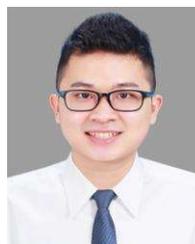

**Viet Quoc Pham** was born in Tay Ninh, Vietnam. He received the B.Eng. degree in Telecommunications Engineering from Ho Chi Minh City University of Technology in 2018. He is currently a postgraduate student at the Faculty of Electrical and Electronics Engineering, Ho Chi Minh City University of Technology. His research interests are the areas of signal processing and wireless communication systems.

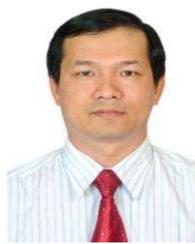

**Ha Hoang Kha** was born in Dong Thap, Vietnam. He received the B.Eng. and M.Eng. degrees from Ho Chi Minh City University of Technology, in 2000 and 2003, respectively, and the Ph.D. degree from the University of New South Wales, Sydney, Australia, in 2009, all in Electrical Engineering and Telecommunications. From 2000 to 2004, he was a research and teaching assistant with the Department of Electrical and Electronics Engineering, Ho Chi Minh City University of Technology. He was a visiting research fellow at the School of Electrical Engineering and Telecommunications, the University of New South Wales, Australia, from 2009 to 2011. He was a postdoctoral research fellow at the Faculty of Engineering and Information Technology, University of Technology Sydney, Australia from 2011 to 2013. He is currently a lecturer at the Faculty of Electrical and Electronics Engineering, Ho Chi Minh City University of Technology, Vietnam. His research interests are in digital signal processing and wireless communications, with a recent emphasis on convex optimization and machine learning techniques in signal processing for wireless communications.

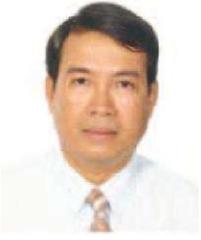**Nguyen Minh Hoang** received the B.Eng. degree from Ho Chi Minh City University of Technology in 1986, M. Eng. degree from Asian Institute of Technology (AIT), Thailand in 1994, and the Ph.D. degree from Institute of Communication Network, Vienna University of Technology, Austria in 2001. He is currently with Saigon Institute of ICT (SaigonICT). His research interests are of computer and communication networks, wireless and mobile communication networks.